\documentclass[sigconf]{acmart}

\title{Heroes and Zeros: Predicting the Impact of New Video Games on Twitch.tv}

\author{Isaac Jones}
\affiliation{%
  \institution{Arizona State University}}
\email{ipjones@asu.edu}

\author{Huan Liu}
\affiliation{%
  \institution{Arizona State University}}
  \email{huan.liu@asu.edu}

\usepackage{amsmath}
\usepackage{graphicx}
\usepackage{algorithm2e}
\newtheorem{hyp}{Hypothesis}

\begin{document}

\begin{abstract}
Video games and the playing thereof have been a fixture of American culture since their introduction in the arcades of the 1980s. However, it was not until the recent proliferation of broadband connections robust and fast enough to handle live video streaming that players of video games have transitioned from a content consumer role to a content producer role. Simultaneously, the rise of social media has revealed how interpersonal connections drive user engagement and interest. In this work, we discuss the recent proliferation of video game streaming, particularly on Twitch.tv, analyze trends and patterns in video game viewing, and develop predictive models for determining if a new game will have substantial impact on the streaming ecosystem.
\end{abstract}

\maketitle



\section{Introduction}

While video games are often considered a useless or wasteful past-time, many have turned that past-time into successful careers and a $22.41$ billion dollar industry\footnote{In $2014$ according to the Entertainment Software Association, the applicable trade association.} has arisen around video games and video gaming. Of this multi-billion dollar industry, ``Gaming Video Content", a secondary market for video content about gaming news, gaming lifestyle, and the games themselves, was estimated to be $\$3.8$ billion in size in $2015$. One site in particular, Twitch.tv, captures approximately $43\%$ of that market\footnote{According to www.superdataresearch.com}.

Twitch.tv started as an offshoot of Justin.tv, which originally streamed the day-to-day life of Justin Khan, one of the four co-founders. The site soon expanded to other streaming subjects, allowing each user to stream whatever they wanted. Streams of users playing video games soon became extremely popular on the site, so much so that Twitch.tv was created, the name inspired by the term ``twitch gaming"\footnote{Twitch gaming refers to a style of gameplay that requires extremely fast reaction times.}.

Twitch.tv became so popular that its userbase eclipsed that of the parent site, aided by a number of events hosted on the site that went ``viral", including a crowdsourced play-through of \textit{Pok\'{e}mon Red} which became known as ``Twitch Plays Pok\'{e}mon" and spawned a number of memes that persist on Twitch's network of content consumers and producers. Some time after this event attracted 6.5 million viewers over the course of 5 days, the original Justin.tv was abruptly shut down. Despite this meteoric rise in popularity, little attention has been paid to the Twitch.tv as either a social network or as a bellwether for success in the gaming market.

Much research on video games has been conducted from the psychological side, studying addictiveness and other social issues on the individual level. Even work in computer science has been focused largely on the individual player level, focusing on player churn, progression, and social habits. Some studies have expanded to analyze small groups of players, but there is no research work on the trends of very large populations of players across games, despite the potential value for content producers, both on Twitch's streaming platform and in the general gaming industry.

In addition, very little attention has been given to the massive numbers of interpersonal connections that arise from Twitch. As we will discuss, normal activity on Twitch forms or reinforces thousands to millions of social bonds every day. However, the possible impacts of these bonds remain unstudied. 

In this work, we will describe the relevant data available directly from Twitch.tv, demonstrate the dynamicity of social relationships through viewership habits on Twitch, analyze how the emergence of new games affects these viewership patterns, and finally develop predictive models to determine which games are likely to affect viewership patterns. In addition, we will discuss some of the related work alluded to above and provide some future directions for other work based on the Twitch platform.

\section{Twitch.tv Data} \label{sec:data}

\begin{figure}
\begin{centering}
\includegraphics[width=0.95\columnwidth]{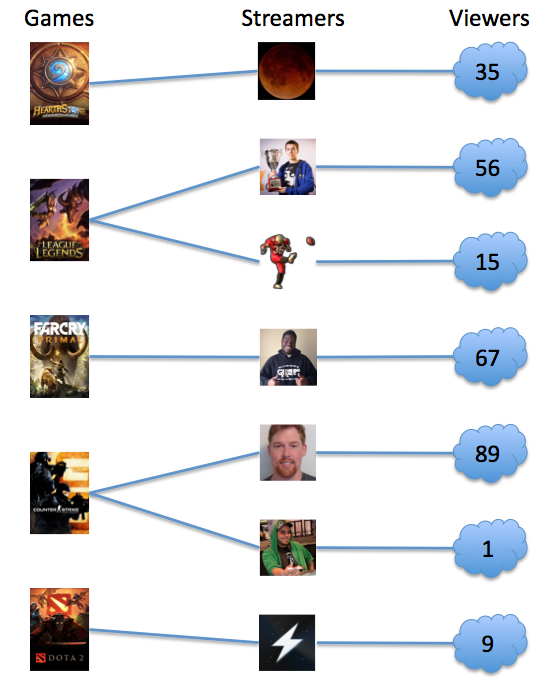}
\caption{A simplified view of the Twitch.tv network depicting the relationships between games, streamers, and viewers.}\label{fig:network}
\end{centering}
\end{figure}

At any given time, Twitch's set of content producers and consumers form a tripartite network, with one-to-many relationships between the layers of the network. The first layer, depicted left-most in Figure~\ref{fig:network}, is \textit{games}. Each game is a unique video game (that is, \textit{Fallout 3}, \textit{DOTA 2}, \textit{Overwatch}, etc.) or one of Twitch's designated topic categories. Topic categories include ``Creative" for creative work, ``Talk Shows" for radio-style talk shows, and ``Programming" for coding work, among others. Topic channels need not be gaming-related. For example, a marathon rebroadcast of Bob Ross's \textit{Joy of Painting} was extremely popular and there is now an ``Eating" category, though most are at least peripherally gaming-related.

The second layer, \textit{channels}, forms a many-to-one relationship with the games layer. Depicted in the middle in Figure~\ref{fig:network}, each channel represents a person or group of people playing a particular game. There is no limit to the number of channels that can play a particular game. Channels are not bound to a particular game for their entire lifetime, but they can only be playing one game at any particular time. Channels are also not obligated to stream at any time or for any length of time, but Twitch has developed a partner system to encourage channels to stream regularly.

The final layer, \textit{viewers}, forms a many-to-one relationship with the channels layer. Transitively, it thus forms a many-to-one relationship with the games layer. Depicted right-most in Figure~\ref{fig:network}, this layer is unique in that individual viewers are not always uniquely identifiable. Twitch allows viewers to be anonymous, though in an anonymous context, they cannot interact with the channel directly. Once viewers sign in on Twitch, they can interact with the channel directly by participating in a channel's chat room. Logged-in users can also directly support the channel by ``following", essentially bookmarking the channel, or ``subscribing" to the channel. Subscribing constitutes direct financial support of the channel, as each subscription costs the viewer \$5 per month. However, only channels that have earned membership in Twitch's Partner Program are eligible for subscriptions.

In order to analyze the relationships between games, streamers, and viewers, we collected snapshots of Twitch's network detailing which games are being played, how many streamers are playing each game, and how many viewers are watching each streamer (and subsequently, each game). During the period spanning April 9th to June 12th of 2016, we collected such snapshots every 15 minutes. Snapshots which did not have complete data were discarded, leaving $5,150$ valid snapshots. Multiple consecutive missing snapshots were observed primarily when Twitch was under heavy load, usually during major eSports tournaments. Among all of these snapshots, we observed $13,951$ unique, identifiable games being played, speaking to the diversity of interest on Twitch.

\begin{figure}
\begin{centering}
\includegraphics[width=0.95\columnwidth]{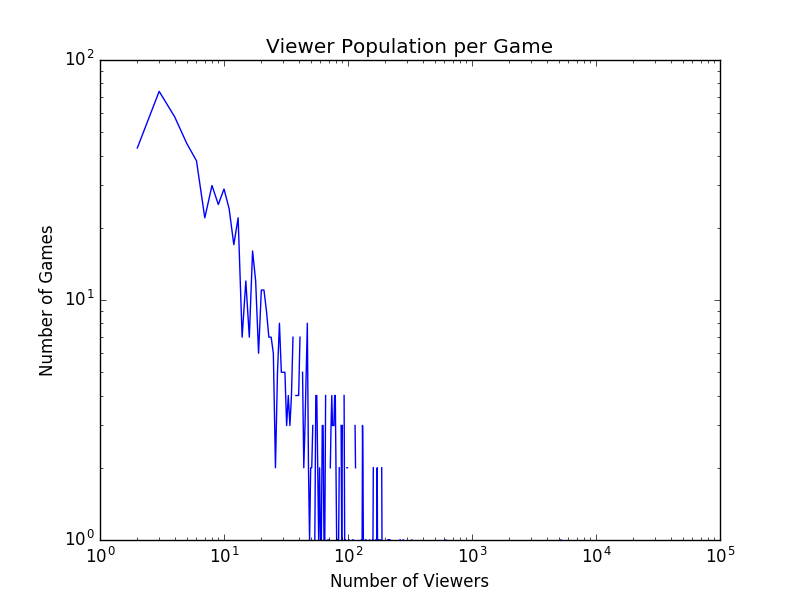}
\caption{Number of games with viewer population vs. number of viewers on April 9th at 6:00 AM GMT. Note that both axes are log-scale, indicating that this trend follows a power-law distribution.}\label{fig:plaw_view}
\end{centering}
\end{figure}

\begin{figure}
\begin{centering}
\includegraphics[width=0.95\columnwidth]{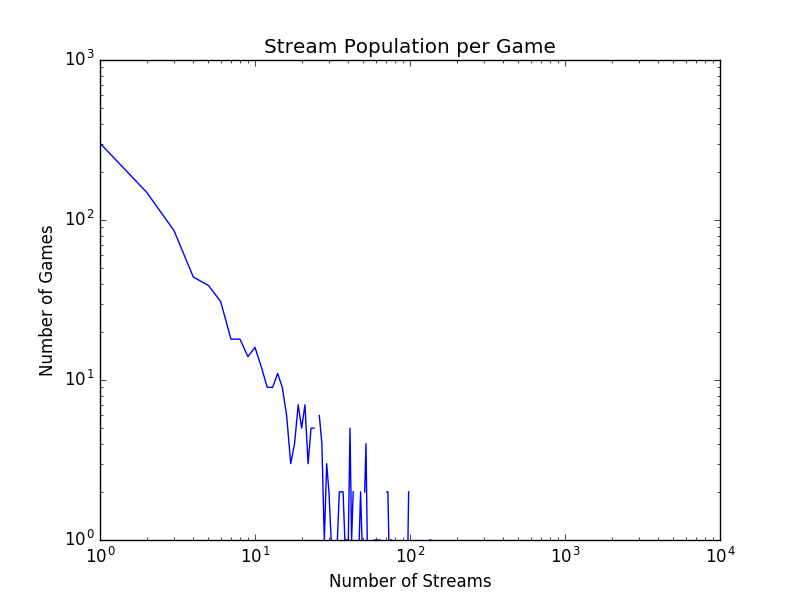}
\caption{Number of games with streamer population vs number of streamers on April 9th at 6:00 AM GMT. Note that both axes are log-scale, indicating that this trend follows a power-law distribution.}\label{fig:plaw_stream}
\end{centering}
\end{figure}

\begin{figure}
\begin{centering}
\includegraphics[width=0.95\columnwidth]{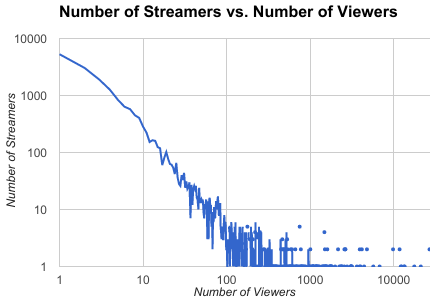}
\caption{Number of viewers vs. number of streams with that viewer population on April 9th at 6:00 AM GMT. Note that both axes are log-scale, indicating that this trend follows a power-law distribution.}
\end{centering}
\end{figure}

In this work, we claim that the set of games, streamers, and viewers can be thought of as a social network. This claim is born out both by the data and by considering the forces that operate on the network. Viewers of game content, who are by far the majority of Twitch's users, do not simply view sterile, homogenous game content. A viewer's experience viewing a particular game is influenced by the streamer's choices. These choices may include a camera showing the streamer's face in a small area of the screen (analogous to picture-in-picture), audio commentary about the game by the streamer, audio interaction with viewers participating in the associated chat channel, and even the style with which the streamer plays the game. Thus, the viewer interacts with the streamer, both explicitly in the chat channel and implicitly in choosing that stream over another playing the same game, and a social link is formed. Conversely, the streamer interacts with viewers through some of the same means, by choosing which game(s) to play, choosing to show or hide his/her face, etc. Thus, social relationships form between streamer and viewer. That viewers willingly and frequently pay to support streamers through the subscription system is a testament to the strength of these bonds.

Quantitatively, Twitch's streaming network also demonstrates many of the same properties as traditional social networks like Facebook and Twitter. For example, both the number of streamers playing a particular game and the number of viewers watching a particular game follow power-law distributions. Similarly, the number of viewers watching each stream of a particular game follows a (weaker) power-law distribution. Graphs of all three of these observations can be seen in Figures~\ref{fig:plaw_view},~\ref{fig:plaw_stream}, and~\ref{fig:plaw_s_v}.

\begin{figure}
\begin{centering}
\includegraphics[width=0.95\columnwidth]{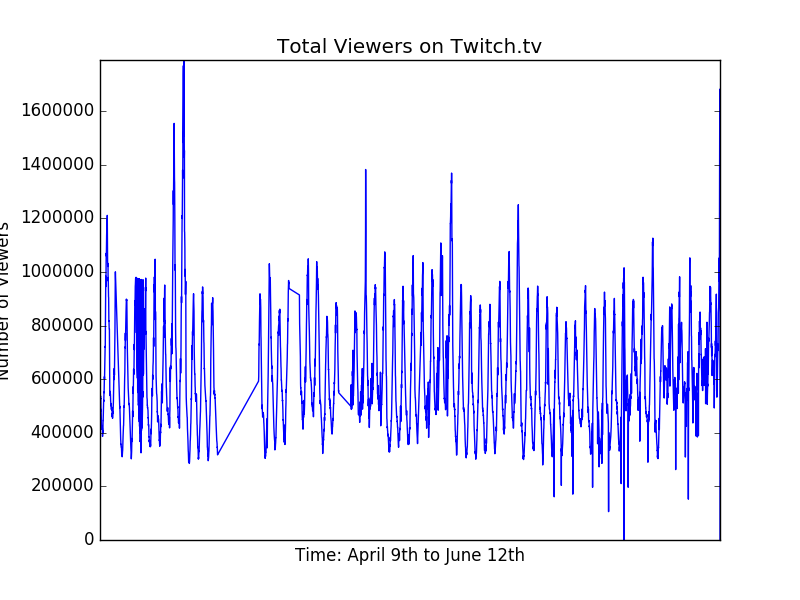}
\caption{Total Viewers on Twitch.tv vs time. This graph shows a clear daily cycle of viewership on Twitch, and also indicates an increase in viewership during the weekends.}\label{fig:viewTime}
\end{centering}
\end{figure}

\begin{figure}
\begin{centering}
\includegraphics[width=0.95\columnwidth]{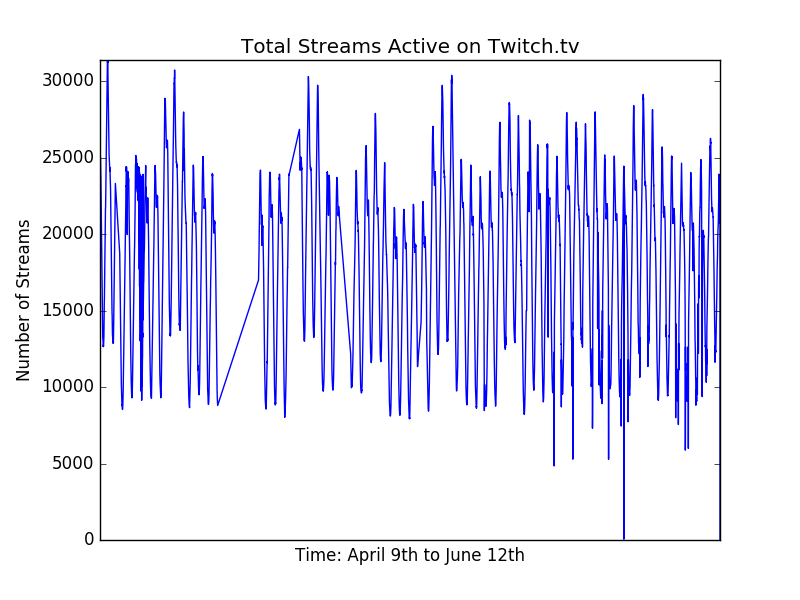}
\caption{Total Streamers on Twitch.tv vs time. Similar to Figure~\ref{fig:viewTime}, this graph shows a strong daily cycle and indicates a weekend boost. Interestingly, the boost appears more noticeable in this graph.}\label{fig:streamTime}
\end{centering}
\end{figure}

We can also expect to see cyclical patterns of viewership and streamership pursuant to the standard daily cycles of human behavior. Since the user base of Twitch.tv is primarily North American or European, we expect to see strong daily viewership cycles corresponding to these time zones. Figures~\ref{fig:viewTime} and~\ref{fig:streamTime} clearly show the daily pattern of both streaming and viewing, respectively, as well as small increases over the weekend, when both streamers and viewers have more time to stream and view. Interestingly, the weekend increase is more noticeable in the streamership graph. This may be because `casual' streamers that do so for entertainment and not to make a living have longer or more frequent streaming sessions during the weekend, while viewership does not change as much over the weekend. More analysis is needed to confirm this observation.

\section{Stream Testing}
\label{sec:streams}

The Twitch network is a dynamic network with many factors affecting the streaming and viewing population at any given time. One of the most apparent factors is the set of games available for streaming and viewing. This set is influenced primarily by the release of new games into the marketplace. Thus, we expect to find that newly released games cause a substantial shift in viewership patterns of the existing games. In order to determine if this is true, we must first determine if there are, in fact, substantial shifts in viewership patterns over time. Logically, if there are no shifts in viewership, new games cannot cause shifts. Phrasing this as a hypothesis, we propose Hypothesis~\ref{hyp:super}, a null hypothesis:

\begin{hyp}
The distribution of a game $g$'s viewership, $v(g)$, does not change over time.\label{hyp:super}
\end{hyp}

Since, as we discussed in the previous section, game viewership on Twitch is highly cyclical, it is not reasonable to analyze Twitch viewership as if it were a stream of continuous values. The data spanning say, $4$:$00$-$5$:$00$ AM would appear very different from the data spanning $4$:$00$-$5$:$00$ PM. Thus, we focus our testing on cycles concomitant with the natural cycles of Twitch viewership, namely cycles with day-long or mutli-day periods. In this analysis, we use cycles of $1$, $2$, $3$, and $7$ days. While single-day and week-long cycles are self-evident, we added $2$ and $3$ day cycles to capture the difference between weekend and weekday viewership patterns. Thus, we revise our previous null hypothesis to that of Hypothesis~\ref{hyp:true}.

\begin{hyp}
The distribution of a game $g$'s viewership, $v(g)$, does not change over $1, 2, 3,$ and $7$ day cycles.\label{hyp:true}
\end{hyp}

In order to test this revised hypothesis, we rely on the technique of~\cite{kifer2004detecting}. This algorithm allows for the successful testing of viewership patterns over rolling, configurable windows. The algorithm for one window is reproduced in Algorithm~\ref{alg:stream}. In our implementation, we consider each game's viewership timeline a separate stream of data, and test each of these streams over all four windows specified above. To account for the invalid snapshots discussed previously, our implementation purges the test window(s) upon encountering an invalid snapshot.

\begin{algorithm}
 \SetAlgoLined
 {\bf Input:} Stream to be tested, $s$; Window size $k$
  $W_1 \leftarrow$ First $k$ values of $s$ \\
  $W_2 \leftarrow$ Next $k$ values of $s$ \\
 \While{$s$ persists} {
 Report change, if applicable. \\
 \uIf{change exists}{
 $W_1 \leftarrow W_2$ \\
 $W_2 \leftarrow$ Next $k$ values of $s$ \\
 }
 \uElse{
 Slide $W_2$ by one value from $s$
 }
 }
\caption{Stream hypothesis testing algorithm from~\cite{kifer2004detecting}}\label{alg:stream}
\end{algorithm}

Algorithm~\ref{alg:stream} does not, however, specify a change detection method. This is intentional on the part of the original authors as it permits the usage of whichever change test is most appropriate. Since our final hypothesis only specifies the subject game and time scale as parameters and we already use those to determine stream composition and window size(s), we require a non-parametric test for change in the stream. For our streams, we chose the Kolmogorov-Smirnov (KS) Test~\cite{massey1951kolmogorov}, as it is a reliable, non-parametric test of distribution difference. Performing the KS Test on our game streams per Algorithm~\ref{alg:stream} resulted in $100,628$ distribution changes detected with $\alpha \leq 0.05$ across the $13,951$ game streams, an average of $7$ per game. Table~\ref{tab:changes} shows the number of distribution changes detected per window size. Henceforth, we will refer to these as \textit{change events}.


\begin{table}
\centering
\begin{tabular}{|c|c|c|c|c|}
\hline
$1$ Day & $2$ Day & $3$ Day & $7$ Day & Total \\
\hline
$ 51,686 $ & $ 27,059$ & $17,235$ & $4,648$ & $100,628$ \\
\hline
\end{tabular}
\caption{Number of significant distribution changes (change events) detected for each window size tested using Algorithm~\ref{alg:stream} and the KS Test~\cite{massey1951kolmogorov}.}\label{tab:changes}
\end{table}

\begin{figure}[h]
\begin{centering}
\includegraphics[width=0.95\columnwidth]{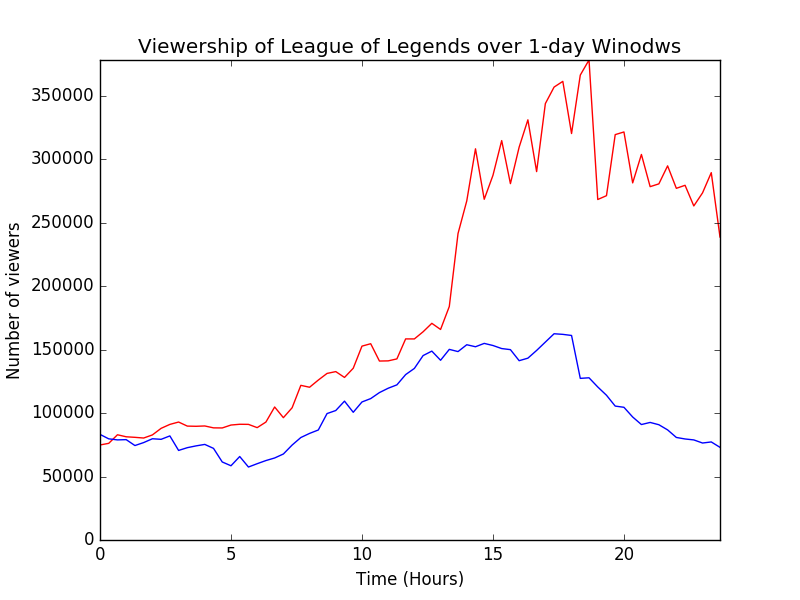}
\caption{\textit{League of Legends} viewer distribution over two 1-day periods. The KS Test detected a significant change between these two distributions.}\label{fig:league_change_1}
\end{centering}
\end{figure}

\begin{figure}
\begin{centering}
\includegraphics[width=0.95\columnwidth]{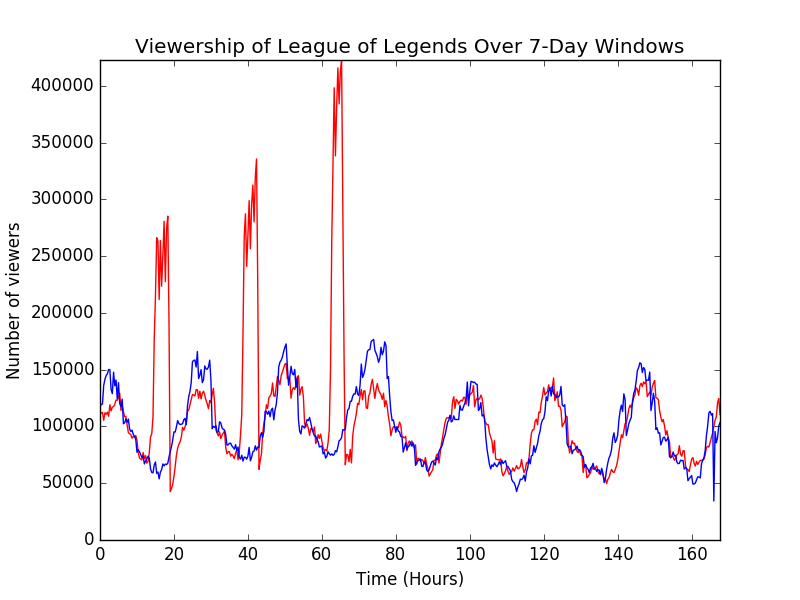}
\caption{\textit{League of Legends} viewer distribution over two 7-day periods. The KS Test detected a significant change between these two distributions.}\label{fig:league_change_2}
\end{centering}
\end{figure}

Such a large number of changes in viewership patterns, even on the longest cycles, indicates that the null hypothesis proposed earlier is throughly rejected. That is, there is detectable variation across all four windows sizes. In order to better understand the changes in viewership distributions, Figures~\ref{fig:league_change_1} and~\ref{fig:league_change_2} show examples of viewer distributions at the time of a change event for the game \textit{League of Legends} over 1 and 7 days. Both of these graphs show a large increase in viewership occurring over a very short period of time. In the case of the second graph, this increase of viewership is repeated three times and then disappears. This pattern in viewership is likely caused by an eSports event occurring for League of Legends, causing many interested viewers to tune in during the competition. Interestingly, this competition is likely a European competition, as it can be clearly seen that the massive spikes occur some hours before the typical daily cycle reaches its peak.

\begin{figure}[h]
\begin{centering}
\includegraphics[width=0.95\columnwidth]{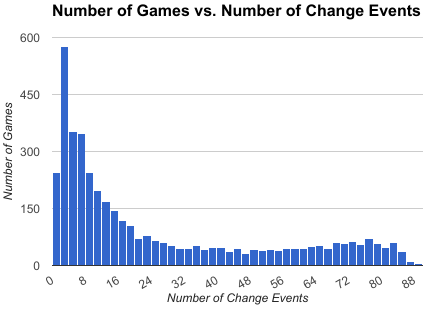}
\caption{Histogram of number of change events and number of games with that number of change events. Note that unlike previous figures, this is not power-law distributed.}\label{fig:gamesChanges}
\end{centering}
\end{figure}

Continuing to inspect the distribution of change events, Table~\ref{tab:game_changes} shows the ten games whose viewership streams have the most associated change events while Figure~\ref{fig:gamesChanges} shows the distribution of the number of changes against the number of games that have those changes. Interestingly, Table~\ref{tab:game_changes} shows that the games that have the most change events are not the most popular games. Many of these games are older games that continue to be popular in the gaming community, but do not command the attention that they once did. For example, \textit{Final Fantasy VII} was released in $1997$ and \textit{Diablo II: Lord of Destruction} was released in $2001$. This may indicate that games like these serve to fill a gap between releases of more popular games, when the newness of newly-released games wears off for streamers and/or viewers and they return to playing and/or watching games that they are familiar with until another new game comes around. More investigation would be needed to confirm this phenomenon, but being the subject of change events both when they lose popularity and then subsequently regain it would explain the high number of change events.

\begin{table}
\centering
\begin{tabular}{|c|c|}
\hline
\textbf{Game Name} & \textbf{Number of Changes} \\
\hline
\textit{Heroes of the Storm} & $89$ \\
\hline
\textit{Final Fantasy VII} & $88$ \\
\hline
\textit{Sid Meier's Civilization V} & $87$ \\
\hline
\textit{Magic: The Gathering} & $87$ \\
\hline
\textit{Street Fighter: V} & $86$ \\
\hline
\textit{StarCraft II} & $86$ \\
\hline
\textit{Diablo II: Lord of Destruction} & $86$ \\
\hline
\textit{The Forest} & $86$ \\
\hline
\textit{Mortal Kombat X} & $86$ \\
\hline
\textit{Final Fantasy X/X-2 HD Remaster} & $86$ \\
\hline
\end{tabular}
\caption{Number of significant distribution changes detected for each of the ten most volatile games.}\label{tab:game_changes}
\end{table}

Interestingly, Figure~\ref{fig:gamesChanges} does not follow the same power-law we would expect to see given that there were power-law relationships between the three slices of the tripartite graph that makes up the Twitch network. It is difficult to see intuitively why this would not also show a power-law relationship, given that some of the same popularity dynamics are at play in the decision to change games as it is to select a game, as demonstrated by Figure~\ref{fig:plaw_view}. More investigation is needed into the precise nature of the migration patterns between games.

Knowing that there are significant fluctuations in game viewership over time, we move on to the next step, determining if these change events are related to the debut of new games on the Twitch network. In the next section, we discuss our method of linking new games with change events on existing game streams, and the value of such a linking in predicting if games debuting in the future will cause similar impacts.

\section{Video Game Debuts}
\label{sec:debuts}

Knowing that fluctuations exist in Twitch viewership distributions naturally leads us to the topic of determining the causal factors of those fluctuations. In the previous section, we speculated that changes in the set of games available to stream would be a large factor in causing fluctuations. We believe this to be the case from our manual observation of the popularity of newer games on Twitch. In addition, this makes intuitive sense: users of social networks often exhibit novelty-seeking behavior, and seeking out and viewing streams of new games fits that behavior pattern.

In order to test this behavior, we analyzed all of our network snapshots in time order and recorded the time at which each of the nearly $14$k games `debuted' on the Twitch network, that is, the first time they appeared in a snapshot. It is important to note here that this debut time does not necessarily match the game's actual release date. Just as movie reviewers often see movies before they are released to general audiences, game reviewers receive review copies of games in advance of the street release date. The popularity of streaming websites and the massive popularity of some of the largest streamers has lead publishers to allow some streamers to play games on stream in advance of their street release date\footnote{goo.gl/3nACrm}. This builds ``hype" for the game and, theoretically, improves sales. However, it does mean that we cannot use a game's release date to determine when that particular game starts being available in the Twitch game pool.

After finding the `debut' time for each game, we counted the number of change events that occurred in the 30 minutes following the game's debut. 30 minutes, we reason, is long enough that viewers interested in the new game will have switched their viewership to the new game, but not so long as to falsely attribute unrelated events to the game of interest. Note that this analysis excludes games that were streamed every day on Twitch, as no games which debuted in the first day of snapshots can possibly have change events associated with their debut, given that the shortest window size is one day. Table~\ref{tab:game_events} counts the number of games with and without change events.

\begin{table}
\centering
\begin{tabular}{|c|c|}
\hline
With Change Events & Without Change Events \\
\hline
$7,174\text{  }(51.4\%)$ & $6,777\text{  }(48.6\%)$ \\
\hline
\end{tabular}
\caption{Number and fraction of game debuts with and without coinciding change events.}\label{tab:game_events}
\end{table}

Interestingly, this analysis indicates that many game releases have a substantial impact on the Twitch viewing population. In fact, this analysis indicates that the majority of game releases are associated with change events. This does not, however, mean that viewership of the most popular games are affected. It is likely that the large population of games with change events indicates that there is heavy competition for the ``tail" of the power-law distribution depicted in Figure~\ref{fig:plaw_view}. This revelation, in combination with the results depicted in Table~\ref{tab:game_changes} and Figure~\ref{fig:gamesChanges}, leads us to believe that a large part of the competition for viewership occurs in the middle ranges of games. Indeed, none of the games in Table~\ref{tab:game_changes} are regularly in the top $10$ games on Twitch. The fact that multiple games debut in the same snapshot (this must occur since there are $5,150$ snapshots and $13,951$ games) supports this theory, as the viewership of the most popular games is not erratic enough to account for these changes. Additionally, a large portion of game debuts linked to change events supports our theory that the set of available games influences streaming habits.

This leads us naturally to the question of prediction. Since game debuts with and without associated change events are approximately balanced, we would like to predict which games will have an impact on Twitch viewership. In order to do this, we supplemented the viewership data from Twitch with data about each game provided by GiantBomb\footnote{http://www.giantbomb.com/api/}. GiantBomb provides an open-access database of metadata concerning individual video games, which we use in the next section to build predictive models.

\section{Success Prediction}
\label{sec:success}

With mappings from game deputs to change events as well as rich data about each individual game, we would like to determine the feasibility of predicting if a game will be successful on Twitch (that is, cause at least one change event) given the data available from a service like GiantBomb. Not all games on Twitch could be matched to games on the GiantBomb service, and those games have been excluded from our analysis from the start. For example, the ``Creative", ``Programming", and ``Talk Shows" pseudo-games do not have entries on GiantBomb, so we excluded them from our analysis. In addition, unresolvable game entries like ``StarCraft II" were excluded, as that could refer to \textit{StarCraft II: Wings of Liberty}, \textit{StarCraft II: Heart of the Swarm}, or \textit{Starcraft II: Legacy of the Void}, the base game and its two expansion packs. Trimming games that could not be supplemented with GiantBomb data removed only $245$ games, less than $2\%$ of the total population\footnote{Resulting in the $13,951$ number cited early. The original set included $14,196$ games and pseudo-games.}.

The data avialable from Giantbomb is rich in nature by virtue of its source, being supplied by enthusiasts and the gaming community. Table~\ref{tab:gb_features} shows the features available from the GiantBomb data. In the classification techniques listed below, we use all features present in a particular game's description. In the case of the description and short description features, we use the length of these descriptions. For features that pertain to characters, objects, people, locations, and reviews, we use the number of such objects. A number of additional time-based features were computed and used, including game age, difference between date added and last updated, date added and release date, original release date and expected release date, and date added and last updated.

\begin{table}[h]
\centering
\begin{tabular}{|c|c|}
\hline
Aliases & Characters \\ 
\hline
Concepts & Developers \\
\hline
Date Added & Date Last Updated \\
\hline
Description & Short Description \\
\hline
Platforms & Publishers \\
\hline
Franchises & Genres \\
\hline
Killed Characters & Debuted Characters \\
\hline
Locations & Debuted Objects \\
\hline
Debuted Locations & Objects  \\
\hline
Debuted Concepts & People \\
\hline
Debuted People & Videos \\
\hline
Main Image & All Images  \\
\hline
User Reviews & Staff Reviews \\
\hline
Expected Release Date & Original Release Date \\
\hline
Themes & Re-Release Dates \\
\hline
Rating & Similar Games \\
\hline
\end{tabular}
\caption{Additional features concerning individual games available from the GiantBomb API.}\label{tab:gb_features}
\end{table}

Using this trimmed data, we used the following three classification algorithms to determine if predicting impactful games is possible and if so, what features were important for prediction.

\begin{itemize}
\item Decision Tree (DT): A decision tree model per~\cite{breiman1984classification}, we included decision trees since the model is easy to learn and highly interpretable. In addition, decision trees easily handle categorical data, which is frequently encountered in video game contexts.

\item Random Forest (RF): Similar to decision trees, random forests per~\cite{breiman2001random} train a large number of randomized decision trees and ensemble the results into a final class decision. Random forests retain a large amount of interpretability since they are simply made up of decision trees while enhancing performance through ensembling. Similar to decision trees, random forests cleanly handle categorical data.

\item One-Class SVM (SVM): This technique, from~\cite{manevitz2001one}, differs from the other two techniques as it is not based on decision trees and thus loses the interpretability of trees. We included this classifier for its outlier detection abilities. Since game releases associated with change events in the most popular games are rare, an outlier detection strategy may provide better results than decision trees.
\end{itemize}

Using $10$-fold cross validation, we trained each of the three models above using the complete data set described above. The Decision Tree and the Random Forest classifiers were limited to a depth of 5 to avoid overfitting. In Table~\ref{tab:class_results}, we report the average of the accuracy, precision, recall, and F1 measure for all three classifiers.

\begin{table}
\centering
\begin{tabular}{|c|c|c|c|c|}
\hline
Model & Accuracy & Precision & Recall & F1 \\
\hline
DT & $0.608$ & $0.582$ & $0.851$ & $0.690$ \\
RF & $0.612$ & $0.595$ & $0.814$ & $0.687$ \\
SVM & $0.555$ & $0.599$ & $0.410$ & $0.486$ \\
\hline
\end{tabular}
\caption{Results from the three classifiers on the task of predicting if a game's debut will be coincident with change events.}\label{tab:class_results}
\end{table}

The results from these classifiers are indicative of the true difficulty of the task. It is also important to note that these classifiers do not represent the best possible performance. Deep learning or other tools may be able to provide better performance. In this work, we seek to demonstrate that performing this classification is possible and valuable. All three classifiers provide better results than random, indicating that it is possible to determine games that will be successful automatically. As expected, the random forest classifier outperformed the decision tree classifier, though the difference in performance is small. Surprising, however, were the results from the one-class SVM classifier. The only metric on which the SVM classifier is competitive is precision, and recall suffers. 

Since one-class SVMs is an outlier detection approach to this problem, as we discussed above, this mode of failure is quite surprising. We would expect this approach to fail in such a way that overgeneralizes, leading to very high recall, but low accuracy and precision. However, the approach provides a competitive precision, but low accuracy and recall. This strange performance may indicate that categorical attributes of the data are more useful in determining the impact of video games on Twitch, as the SVM model and tree-based models necessarily deal with these attributes differently.

Inspecting the decision trees generated from our classification efforts shows that the length of game's text description from GiantBomb is the most important feature, counter to our expectation from the difference in performance between decision trees and the SVMs classifiers. However, we do not believe that the dominance of this feature is surprising. As GiantBomb's data is supplied by the gaming community, a longer description may indicate more enthusiasm about the game from the community. This enthusiasm translates to a higher willingness to watch the game when it is available on Twitch, which translates to a higher likelihood of impact events. Similarly, the dominance of this feature is reassuring for the long-term usefulness of our models. Since there is so much media coverage of games before their release dates, 

After this feature, categorical features dominate, which matches our expectations. The most heavily used among them are the platforms on which the game is available, with preference given to the latest generation of consoles. This is reasonable, as newly released games, which are naturally more likely to receive interest, are likely to be released on the most recent platforms. In addition, the newest platforms have streaming capability built directly in to the software making it easier for the newest games to be streamed. In fact, Sony's Playstation 4 can be configured to begin streaming whatever game is being played at the press of a single button.

Somewhat surprising in these results is the poor gain in performance from the Random Forest model, which only slightly outperforms Decisions Trees in accuracy and underperforms in the F1 measure. This lack of substantial change in performance indicates that the features discovered by the decision trees are the most useful features, and subsequent features do not hold a substantial degree of predictive value. To test this claim, Table~\ref{tab:results_new} shows the results of training the Decision Tree and Random Forest classifier without the top feature, the length of the game's long description.

\begin{table}
\centering
\begin{tabular}{|c|c|c|c|c|}
\hline
Model & Accuracy & Precision & Recall & F1 \\
\hline
DT & $0.600$ & $0.575$ & $0.840$ & $0.683$ \\
RF & $0.603$ & $0.581$ & $0.818$ & $0.679$ \\
\hline
\end{tabular}
\caption{Results from the decision tree-based classifiers after removing the most predictive feature, the length of the long description.}\label{tab:results_new}
\end{table}

From this change in the dataset, the difference in performance is apparent, but minor. More interesting than the change in performance is the change in most predictive feature, however. This shifted from the length of the game's description to the number of user reviews. Arguably, these two features actually measure the same thing, as they can both be considered a proxy for community enthusiasm regarding the game, though user reviews will not be available before the game is released while the game's description will be. Interestingly, this second feature does not appear in the first set of decision trees, which indicates that the two features have discriminative power over the same two groups of games.

\section{Related Work}
\label{sec:rel_work}

Though sparse, the body of work related to this research informs the analysis we conducted, and we divide it into two parts for the purposes of discussion. Games research covers video games specifically and migration research covers the patterns and trends by which users transition from one game or site to another.

\subsection{Games Research}

Much of the existing research on video games has been done on a very small set of games. \textit{EverQuest 2}'s player base was studied by analyzing group success~\cite{ahma2011}, success of guilds (in-game player organizations)~\cite{seay2004}, individual player performance~\cite{shim2011mentor}, player enjoyment~\cite{shim2011eq2} and guild lifecycles~\cite{thurau2010}. Player engagement was also studied in \textit{World of Warcraft}, linking group play to decreased leveling progress and guild membership to an increase in hours played~\cite{duch2006}. Categorization of game players into archetypes is a common topic as well, with work considering players of text-based Multi-User Dungeons~\cite{bart1996}, \textit{World of Warcraft}~\cite{bell2013}, and \textit{Star Wars: Galaxies}~\cite{duch2004}. In addition to the characterization of players, Bartle's work~\cite{bart1996} gives some rudimentary recommendations for balancing virtual worlds to maximize enjoyment for all players and~\cite{tayl2015} discusses interactions between play styles from a player perspective. Economic analysis of virtual worlds has also been of interest, particularly in determining if they serve as effective proxies for real-world economies~\cite{castronova2008synthetic}. In particular, \textit{The Kingdom of Loathing} has received scrutiny~\cite{saff2011} as well as \textit{EVE Online}~\cite{tayl2015}. Even the habits of game-playing on mobile devices has started to receive attention, particularly in South Korea~\cite{seok2015}. The sheer quantity of games played by users of the Steam platform\footnote{store.steampowered.com} was investigated~\cite{sifa2015large} and they also found evidence of power-law relationships in game playing times, though they did not investigate number of players playing a game as we have.

\subsection{Migration Research}

The extreme mobility of users between social networks has also inspired research into the migration of users from one social network to another~\cite{kumar2011understanding}. Some researchers have developed techniques to match users across social networks based on usernames~\cite{zafa2013}, behavioural modeling~\cite{liu2014}, and network alignment~\cite{zhang2015}. Viewers and streamers on Twitch have similarly high mobility, though our work benefits from the internal nature of that mobility. Players leaving their current game and switching to a new one, called churn, is the subject of much research. Churn has been studied in \textit{World of Warcraft}~\cite{debe2011}, \textit{Everquest 2}~\cite{shim2010}~\cite{kawale2009}, \textit{EVE Online}~\cite{feng2007}, and one study across $5$ different games~\cite{bauckhage2012}. Our work in this area stands out both for its scale (covering $13,951$ games) and links across games, which other studies lack.


\section{Discussion and Conclusions}
\label{sec:conc}

In this work, we have discussed Twitch.tv and the data available on this platform in detail, the first such discussion available, demonstrated that viewership patterns vary over even short time spans, linked those variations to new games debuting on the Twitch network, and demonstrated that even simple classification algorithms have predictive power on a game's impact, again the first such work. In doing so, we have contributed to the understanding of the complex interaction between video game players and the games they play as well as the games they choose not play, as well as with other players of the same and different games. This understanding could be used to develop games that are more satisfactory for players and viewers alike, and thus are more successful. In addition, it could be used as a recommender system for streamers, who, in growing their following, want to balance enjoyment of their own play time with playing games that are new and interesting.

Future extensions of this work include refining the predictive models for game impact, enhancing churn analysis with the knowledge of not only when streamers switch games but their destination, determining the success factors for streamers in addition to those for games, and analysis of the economics of streaming with respect to the Twitch Partner Program. Refining the predictive models is of particular interest, as our approach currently cannot distinguish between the impact of two games that appear in the same snapshot.

In addition, we identified a number of areas in the paper that warrant further study, mostly around change events. The frequency of change events associated with games in the middle range of popularity warrants further investigative work and such work may benefit from a longer-duration study. In addition, it is worth investigating if other basic social principles, like homophily, apply to the Twitch network as they apply to other social networks.


\bibliographystyle{ACM-Reference-Format}
\bibliography{games_bib}
\end{document}